\documentstyle[12pt,aasms4]{article}

\newcommand {\ie} {{\it  i.e.}}

\newcommand {\eg} {{\it e.g.}}
\newcommand {\etal} {{\it et~al.}}

\newcommand {\go} {\mathrel{\hbox{\rlap{\lower.55ex \hbox {$\sim$}}
        \kern-.3em \raise.4ex \hbox{$>$}}}}
\newcommand {\lo} {\mathrel{\hbox{\rlap{\lower.55ex \hbox {$\sim$}}
        \kern-.3em \raise.4ex \hbox{$<$}}}}    
\def\half{\ifmmode {1\over 2} \else ${1\over 2}$\fi} 
\def\tor{\ifmmode {3\over r} \else ${3\over r}$\fi} 
\def\dwds{\ifmmode \partial W/\partial \Sigma \else $\partial W/
\partial \Sigma$\fi} 
\def\SigG{\ifmmode \Sigma_{GPD} \else $\Sigma_{GPD}$\fi} 
\def\SigR{\ifmmode \Sigma_{RPD} \else $\Sigma_{RPD}$\fi} 
\def\Msun{\ifmmode M_\odot \else $M_\odot$\fi} 
\def\Mdot{\ifmmode \dot M \else $\dot M$\fi} 
\def\rms{\ifmmode r_{ms} \else $r_{ms}$\fi} 
\def\rin{\ifmmode r_{in} \else $r_{in}$\fi} 
\def\kes{\ifmmode \kappa_{es} \else $\kappa_{es}$\fi}
\def\fK{\ifmmode \nu_K \else $\nu_K$\fi} 

\begin{document}

\title{VISCOUS STABILITY OF RELATIVISTIC KEPLERIAN ACCRETION DISKS}
\author{Pranab Ghosh$^{1,2}$}
\affil{Laboratory for High Energy astrophysics}
\affil{NASA Goddard Space Flight Center, Greenbelt, MD 20771}

\lefthead{Ghosh}
\righthead{Stability of Relativsitic Disks}

\vskip 2 truein

$^1$ Senior NAS/NRC Resident Research Associate
 
$^2$ On sabbatical leave from Tata Institute of Fundamental 
Research, Bombay 400 005, India

\vskip 1.2 truein

\centerline{Received July 20, 1998;\quad accepted August 19, 1998}

\vskip 0.25 truein

\centerline{To appear in the Astrophysical Journal Letters}

\vskip 2 truein

\begin{abstract}

We investigate the viscous stability of thin, Keplerian accretion
disks in regions where general relativistic (GR) effects are 
essential. For gas pressure dominated (GPD) disks, we show that
the Newtonian conclusion that such disks are viscously stable is
reversed by GR modifications in the behaviors of viscous stress
and surface density over a significantly large annular region not 
far from the innermost stable orbit at $r=\rms$. For 
slowly-rotating central objects, this region spans a range of 
radii $14\lo r\lo 19$ in units of the central object's mass $M$. 
For radiation pressure dominated (RPD) disks, the Newtonian 
conclusion that they are viscously unstable remains valid after 
including the above GR modifications, except in a very small 
annulus around $r\approx 14M$, which has a negligible influence. 
Inclusion of the stabilizing effect of the mass-inflow through 
the disk's inner edge via a GR analogue of Roche-lobe overflow 
adds a small, stable region around \rms~for RPD disks, but 
leaves GPD disks unchanged. We mention possible astrophysical 
relevance of these results, particularly to the high-frequency 
X-ray variabilities observed by the $Rossi$ $X-ray$ $Timing$ 
$Explorer$.
    
\end{abstract}


\section{Introduction}

Extensive studies of the stability of geometrically thin, optically
thick, Keplerian accretion disks in the Newtonian regime (Lightman 
\& Eardley 1974, henceforth LE; Lightman 1974a,b, henceforth 
L74a,b; Shakura \& Sunyaev 1976; Piran 1978, henceforth P78) have 
established that such disks are unstable to both viscous and 
thermal modes when they are radiation pressure dominated (RPD), but 
not when they are gas pressure dominated (GPD). In this 
{\it Letter}, we show that GPD disks are viscously unstable in an 
essentially relativistic region not far from the innermost stable 
orbit at $r=\rms$, due to general relativistic (GR) modifications 
in the behaviors of the disk's viscous stress and surface density. 
For a slowly-rotating central accreting object, this region spans 
a range of radii $14\lo r\lo 19$ in units of the central object's 
mass $M$ (we use the convention $G = c = 1$ throughout this work). 
We demonstrate that, after including these GR modifications, RPD 
disks remain viscously unstable in this region, except in an 
extremely small annulus around $r\approx 14M$, which has a 
negligible effect. Inclusion of the stabilizing effect of the 
mass-inflow through the disk's inner edge via a GR analogue of 
Roche-lobe overflow (Abramowicz 1981, 1985, henceforth A81, A85) 
adds a further, small, stable annulus around \rms~for RPD disks, 
but leaves GPD disks unchanged. The inner parts of GPD disks 
(around black holes or sufficiently compact, weakly-magnetized 
neutron stars) extending to radii $\go\rms$ are thus expected to 
have blobs of matter: we indicate the relevance of this result 
for high-frequency X-ray variability observed with the $Rossi$ 
$X-ray$ $Timing$ $Explorer$ ($RXTE$).
 
\section{The Viscous Instability}

Viscous stability in thin, Keplerian disks in GR is described by 
the evolution of the disk's surface density $\Sigma (r,t)$, 
which is obtained by combining the conditions of conservation 
of mass and angular momentum. In Kerr metric, these are
$${\cal D}^{\half}{\partial\over\partial r}\left({\cal D}^{-\half}
\Mdot\right) = 2\pi r {\cal A}^{\half}{\cal D}^{-\half}{\partial
\Sigma\over\partial t}\,,\eqno(1)$$
$$\Mdot{\partial\ell\over\partial r} = {\partial\over\partial r}
(2\pi r^2{\cal BC}^{-1}{\cal D}W)\,.\eqno(2)$$
  Here, $\Mdot (r,t)$ is the local, instantaneous mass flux,
$W$ is the vertically-integrated viscous stress, $\ell$ is the 
specific angular momentum, and $\cal A$, $\cal B$, $\cal C$,
$\cal D$ are relativistic factors (Novikov \& Thorne 1973, 
henceforth NT73) whose values reduce to unity in 
the Newtonian regime:
$${\cal A}\equiv 1+{a^2\over r^2}+{2a^2\over r^3},\quad 
  {\cal B}\equiv 1+{a\over r^{3/2}},\quad
  {\cal C}\equiv 1-{3\over r}+{2a\over r^{3/2}}, \quad
  {\cal D}\equiv 1-{2\over r}+{a^2\over r^2}\,.\eqno(3)$$
Here $a$ is the specific angular momentum of the central accreting 
object, and we express lengths and times in units of $M$ in all 
relativistic factors {\it only\/}. 
Equations (1) and (2) yield    
$${\cal A}^{\half}{\cal D}^{-\half}{\partial\Sigma\over\partial 
t} = {\cal D}^{\half}{1\over r}{\partial\over\partial r}\left[
{{\cal D}^{-\half}\over (\partial\ell/\partial r)}
{\partial\over\partial r}\left({\cal BC}^{-1}{\cal D}r^2W\right)
\right]\,,\eqno(4)$$
showing how the viscous stress $W$ determines the evolution of 
$\Sigma$. 

The Newtonian limits of equations (1)-(4) were used by LE and L74b 
to argue that viscous instability sets in when viscous stress 
is a decreasing function of surface density, \ie, $\dwds < 0$, 
while the disk is stable when $\dwds > 0$ (also see 
Pringle 1981). This is transparent upon expressing 
the radial derivative of $W$ in equation (4) in terms of 
\dwds~(L74b), and comparing with the standard form for diffusion 
equations: the effective diffusion coefficient is negative in the 
former case, indicating the formation of density clumps, while it
is positive in the latter case, indicating normal viscous diffusion.
The physical reason is equally transparent: in the former case, 
a local upward fluctuation in $\Sigma$ causes a decrease in the 
viscous stress, making the viscous smoothing of the density 
clump more difficult, and the situation runs away (albeit on the 
slow, viscous timescale). The latter case is stable, as any 
tendency towards clumping is nullified by the stronger viscous 
smoothing it generates. Since for ``standard'' Newtonian Keplerian 
accretion disks (Shakura \& Sunyaev 1973, henceforth SS73) 
$W\propto\Sigma^{-1}$ in RPD regions (LE) and $W\propto
\Sigma^{5/2}$ in GPD regions (see \S 3.2), LE and L74b derived 
the well-known Newtonian result that the former are viscously 
unstable, while the latter are stable.

The crucial point about relativistic disks is that the basic 
criterion for viscous stability in terms of \dwds~remains the 
same as above, as an inspection of equation (4) readily shows. 
This is as expected, since the physical picture given above 
remains unaltered in the GR regime. Therefore, the stability of a 
disk model in the GR regime is determined by the relation between 
$W$ and $\Sigma$ in that regime. In \S 3, we summarize this 
relation for thin, Keplerian disks in the GR regime, and its 
effects on viscous stability. In so doing, we use the GR 
formulation of geometrically thin, optically thick, Keplerian 
accretion disks of NT73 throughout. The formulation of SS73 was 
basically Newtonian, with an appropriate GR boundary condition 
applied {\it ad hoc\/} to the viscous stress at \rms. While the 
consistent GR formulation of NT73 is more suitable for our 
purposes, we indicate later that the effects described here can
be seen qualitatively even in the SS73 formulation, and in the
pseudo-Newtonian generalization (Abramowicz \etal~1988, 
henceforth ACLS) thereof.   
       
\section{Stability in the Relativistic Regime}

The vertically-integrated viscous stress $W$ is obtained in the 
GR formulation (NT73) by combining the conservation laws 
for angular momentum and energy, subject to the boundary condition
that $W$ vanishes at $r=\rms$, where the gas ``falls out'' of the 
disk and spirals down to the central object:
$$W = {\Mdot\over 2\pi}\left({M\over r^3}\right)^{\half}
{{\cal C}^{\half}{\cal Q}\over {\cal BD}}\,.\eqno(5)$$
Here, $\cal Q$ is a relativistic factor (NT73) which 
describes the effects of the boundary conditions on $W$: at 
$r=\rms$, $\cal Q$ vanishes, while in the Newtonian regime, 
$\cal Q$ reduces to unity. $\cal Q$ has been calculated 
for the Kerr metric in Page \& Thorne (1974). Explicit
computations given in this {\it Letter} will be confined to the
Schwarzschild metric\footnote{Extension of our results to 
non-zero values of the angular momentum parameter $a$ presents 
no difficulties or surprises, and will be described elsewhere.}
, in which limit $W$ is given by\footnote
{In both the SS73 formulation and its ACLS generalization, the 
logarithmic term within the square brackets in eq.[6] is 
missing; in addition, the relativistic factor 
$(1-{2\over r})^{-1}$ is missing as well in the SS73
formulation. However, $W$ still has the same qualitative shape.}  
$$W = {\Mdot\over 2\pi}\left({M\over r^3}\right)^{\half}
(1-{2\over r})^{-1}\left[1-\sqrt{6\over r}+\half\sqrt{\tor}
\ln\left({1+\sqrt{\tor}\over 1-\sqrt{\tor}}.{\sqrt{2}-1\over 
\sqrt{2}+1}\right)\right]\,,\eqno(6)$$
and shown in Figure 1. From the Newtonian limit $W\propto 
r^{-3/2}$ at large distances, $W$ passes through a maximum at 
$r\approx 14M$, vanishing at $r=\rms=6M$.

\subsection{Vertical Structure}

The disk's surface density $\Sigma$ is determined by 
its vertical structure, the physics of which is threefold:
vertical hydrostatic equilibrium, vertical radiative transport of
energy, and equation of state (SS73; NT73). In GR, the first is
given by  
$${\partial p\over\partial z} = -{\rho Mz\over r^3}{\cal KC}^{-1}
\,,\eqno(7)$$
where $p$ and $\rho$ are the disk pressure and density, 
respectively, and the relativistic factor $\cal K$ is given by
$${\cal K} = 1-{4a\over r^{3/2}}+{3a^2\over r^2}\,.\eqno(8)$$
Note that this is the improved version of the original NT73 
formulation as given by Riffert \& Herold (1995); for more 
detailed GR formulation and discussion, see Abramowicz, Lanza \& 
Percival (1997) and references therein. The second is given by
$$F = {3\over 4}\left({M\over r^3}\right)^{\half}W{\cal DC}^{-1}
= {2bT^4\over 3\kes\Sigma}\,,\eqno(9)$$
where $F$ is flux of radiant energy from each face of the disk, 
$T$ is the disk temperature, $b$ is the radiation constant, and 
\kes~is the electron-scattering opacity (appropriate for the 
inner disk). The equation of state is given by
$$p = \cases{{\Sigma kT\over hm_p}\, &GPD;\cr
 {1\over 3}bT^4\, &RPD. \cr}\eqno(10)$$
Here, $h$ is the semi-thickness of the disk, $m_p$ is the proton 
mass, and $k$ is Boltzmann's constant.

\subsection{$W-\Sigma$ Relation and Stability}

Together with the relation $W=2\alpha hp$ for the viscous stress, 
$\alpha$ being the standard disk viscosity parameter (SS73; NT73), 
equations (7)-(10) yield the following results for the surface 
densities of GPD and RPD disks:
$$\SigG = \left({m_p\over k\alpha}\right)^{4/5}\left({b\over 
18\kes}\right)^{1/5}M^{-1/10}(W\sqrt{r})^{3/5}
{\cal C}^{1/5}{\cal D}^{-1/5}\,.\eqno(11)$$
 
$$\SigR = \left({16\over 9\alpha\kes^2}\right)W^{-1}
{\cal CKD}^{-2}\,.\eqno(12)$$   
Radial variations of \SigG~and \SigR~are shown\footnote
{The apparent rise in Fig.1 of \SigR~to large values at large 
$r$, and also as $r\rightarrow\rms$, are unphysical, of course.
In reality, the former rise is terminated where the RPD region 
joins smoothly to a GPD region at sufficiently large radii, and 
the latter is terminated close to \rms, where there is a
transition from Keplerian flow to free-fall via a GR 
analogue of Roche-lobe overflow (see text).} in Figure 1 in 
the Schwarzschild limit. $W-\Sigma$ relations for GPD and 
RPD disks are shown explicitly in Figures 2 and 3, respectively. 

For GPD disks, the Newtonian property $\dwds > 0$ is valid for
all radii larger than that at which \SigG~attains its maximum,
\ie, $r\approx 19M$. However, between the 
maxima of \SigG~and $W$, \ie, in the region $14M\lo r\lo 
19M$, $\dwds < 0$, since $W$ decreases with increasing $r$
there, but \SigG~increases. At $r\lo 14M$, $\dwds > 0$ again,
as both $W$ and \SigG~increase with $r$. Thus, between the 
lower, Newtonian branch of the $W-\Sigma$ relation shown in 
Figure 2 (on which $W\propto r^{-3/2}$, and $\SigG\propto 
r^{-3/5}$ as eq.[11] shows, so that $W\propto\Sigma^{5/2}$),  
and the upper, GR branch, both of which are viscously stable, 
there is a considerable transition region ($14M\lo r\lo 19M$ 
in the Schwarzschild limit) which is unstable 
to the viscous mode. Hence, blobs of matter are expected to 
form in this annular region of GPD disks. The ultimate fate 
of these blobs merit further study: it is clear that, as they
drift radially into the stable region at $6M\lo r \lo 14M$,
viscous stresses will tend to smoothen them out, but a 
significant fraction of blobs may survive to the disk's inner 
edge, since the drift timescale through this stable region 
exceeds the fastest stabilization timescale only by a factor 
$\lo 3$.

For RPD disks, the lower, Newtonian branch of the $W-\Sigma$ 
relation (on which $W\propto\Sigma^{-1}$ and $\dwds < 0$) 
is almost identical to the upper, GR branch, with no transition 
region visible in the main panel of Figure 3. A closer inspection 
of the curve's tip (inset) does reveal an extremely small 
transition region ($13.89\lo r\lo 14.26$) over which $\dwds > 0$, 
corresponding to the region between the maximum of $W$ and the 
minimum of $\SigR$ in Figure 1. 
Although this tiny region is viscously stable in
principle, it has no significant effect on the stability 
properties of RPD disks, which would be expected to form blobs 
over their entire extent from the arguments given so far 
(but see below).

Finally, the stabilizing effect (A81, A85) of the mass-inflow 
through the inner edge of the disk to the central object via a 
GR analogue of Roche-lobe overflow (Abramowicz, Calvani \& 
Nobili 1980, henceforth ACN, and references therein) is likely 
to have minor consequences for most of the results found here, 
particularly for GPD disks. Found originally (A81) by applying 
P78's Newtonian stability analysis to the above GR mass-inflow 
through inner edges of thick, RPD, non-Keplerian disks with 
super-Eddington accretion rates (ACN), this effect is now 
understood to stabilize the {\it transonic\/} part (A85) of 
the accretion flow thermally (and possibly also viscously),
as well as the supersonic part of the flow interior to \rms.
However, the radial extent ${\delta r\over\rms}$ of the 
transonic regions at the inner edges of the thin, Keplerian 
disks under consideration here is very small, typically 
$\lo 10^{-2}$ (Miller, Lamb \& Psaltis 1998, henceforth MLP,
and references therein), comparable to the small vertical 
thicknesses (A85) of the transonic flows. For RPD disks, 
this will lead to the formation of an additional, small, 
stable annulus around \rms, but there is virtually no change 
for GPD disks, since this region is stable anyway. Similar
arguments apply to those Keplerian disks around 
weakly-magnetized neutron stars which are terminated at radii
$\go\rms$ by non-gravitational forces, \eg, radiation drag
(MLP).              
            
\section{Discussion}

Underlying the phenomenon described here is a basic difference 
between the scalings of \SigG~and \SigR~with $W$ in 
$\alpha$-disks which does not seem to have been fully appreciated 
so far. Apart from the (minor) effects of the factors 
$\cal C$, $\cal D$ and $\cal K$,  the $W-\Sigma$ relation 
for RPD disks, $\SigR\propto W^{-1}$, is independent of $r$, 
but that for GPD disks, $\SigG\propto (W\sqrt{r})^{3/5}$, is
not. Hence, when $W$ develops a maximum at $r\sim 14M$ as a 
result of the GR boundary condition, and runs through the same 
set of values (see Figure 1) for $r > 14M$ (Newtonian branch) and 
$R < 14M$ (GR branch), corresponding values of \SigR~are 
essentially the same on the two branches, but those of 
\SigG~are quite different (see Figures 2 and 3). The maximum of
\SigG~occurs at that of $W\sqrt{r}$, which is quite different 
from that of $W$. In the region between the two maxima, the 
viscous stress is clearly always a decreasing function of \SigG, 
and the disk viscously unstable. This behavior is robust (but 
see below) in the sense that it depends not on the details of
how GR is introduced, but only on the basic scaling $\SigG\sim 
(W\sqrt{r})^{3/5}$, which is valid even in the Newtonian 
limit: this is why the SS73 formulation and its ACLS 
generalization give results which are qualitatively the same 
as that given here. 

All $\alpha$-disks show the above scalings for RPD and
GPD regions because \SigR~scales as $W^{-1}$ in such disks, but
\SigG~as $(W\Omega^{-1/3})^{3/5}$, $\Omega$ being the 
Keplerian angular velocity: the reason for this is clear from
the vertical structure relations. These relations
involve $\Omega$ twice: in that of vertical equilibrium 
(eq.[7]), and in that of viscous dissipation rate (eq.[9]). 
There is a cancellation between the two $\Omega$s for RPD 
disks, but {\it not\/} for GPD disks, because pressure is 
independent of $\Sigma$ for the former, but not for the latter
(LE; L74b). Of course, this reasoning applies only to 
$\alpha$-disks, as did that of LE. Investigations of the 
corresponding properties of GR non-Keplerian disks, 
particularly advection-dominated disks (Gammie \& Popham 1998
and references therein) appear worthwhile.

Astrophysical applications of our results will be given in 
detail elsewhere. While the focus of the stability analyses 
of inner disks in the 1970s was on bright, galactic black-hole 
candidates, where accretion rates are normally high enough to 
ensure RPD inner regions, it is now clear that accretion rates 
in many X-ray binaries containing black holes and 
weakly-magnetized neutron stars can be low enough ($\Mdot\lo 
10^{-2}$ of the Eddington rate) that inner disks are GPD  
all the way to their inner edges at $\sim\rms$. 
Low-mass X-ray binaries (LMXB), and, in particular, ``atoll'' 
type LMXBs with very low neutron-star magnetic fields and 
accretion rates of the above order, are prime examples of the
latter case. Signatures of the phenomena described here may be 
observable in timing studies of LMXBs by $RXTE$, particularly 
in the kilohertz quasiperiodic oscillations (henceforth kHzQPO; 
see, \eg, van der Klis 1998, henceforth K98). 
We note that the Keplerian frequencies corresponding to 
the unstable region in GPD disks around slowly-rotating central 
objects are $\sim 210-350(M/2\Msun)^{-1}$ Hz, similar to the 
frequency difference between the twin kHzQPOs observed in LMXBs. 
Furthermore, kHzQPO diagnostics of any residual blobs present 
in the region between the disk's inner edge and that of the 
viscously unstable region (see \S 3.2) does not appear 
impossible (Angelini 1998, private communication).    
           
\acknowledgements

It is a pleasure to thank L. Angelini, J. K. Cannizzo, J. H.
Swank, and N. E. White for stimulating discussions, and the 
referee, M. C. Miller, for helpful comments.

\clearpage

\begin{figure}

\centerline{\bf FIGURE CAPTIONS}

\caption {Radial variation of the integrated viscous stress $W$ 
(solid line) in a thin, Keplerian disk in the Schwarzschild 
limit, as well as that of the surface density \SigG~of a GPD 
disk (dashed line), and that of the surface density \SigR~of a 
RPD disk (dotted line), also in the same limit. Radii shown in
units of the central object's mass $M$. Since only the shapes
of the curves matter for this work, vertical units have been 
adjusted for convenient simultaneous display of the profiles. 
Also shown at the top of the panel is the range of radii over 
which GPD disks are viscously unstable, and, at the bottom of 
the panel, the (extremely small) region over which RPD disks are 
viscously stable, because of GR effects (see text).}     

\caption {Integrated viscous stress $W$ vs. surface density 
\SigG~of a Keplerian GPD disk in the Schwarzschild limit. The
lower branch of the curve is the Newtonian one, and the upper
branch the ultrarelativistic one, both of which are viscously
stable (see text). Between these, there is a transition region
(marked by X's) which is viscously unstable (see text), 
corresponding to the range of radii similarly marked at the 
top of Fig.1. Note that the radial coordinate decreases 
monotonically along the curve in going from the lower to the
upper branch.}

\caption {Similar to Fig.2, but for RPD disks. The main panel
shows that the lower, Newtonian branch is almost identical to
the upper, relativsitic branch, both being viscously unstable
(see text), and that there is almost no transition region. 
The inset, a magnification of the tip of the main curve, 
reveals an extremely small transition region which is 
viscously stable, corresponding to the marks at the bottom
of Fig.1, but this has little effect on the overall stability 
of RPD disks.}

\end{figure}

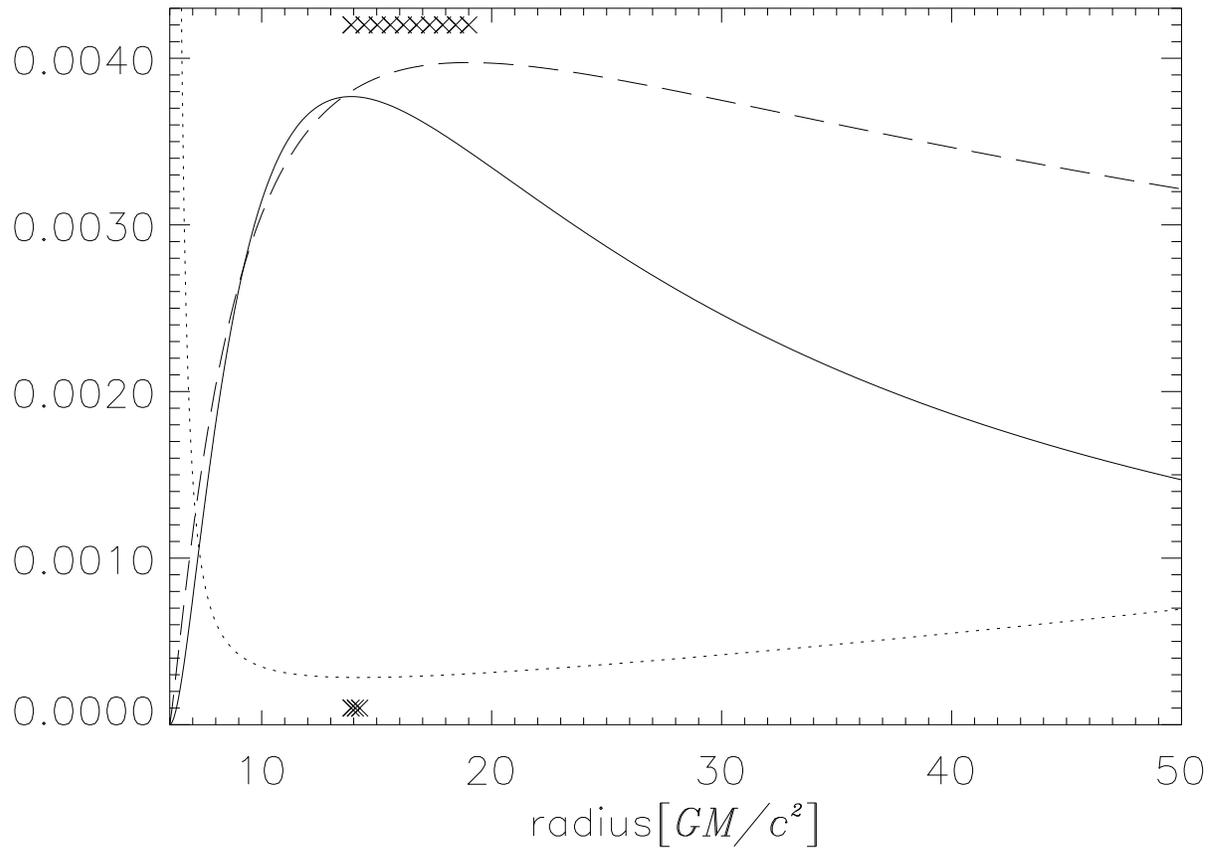
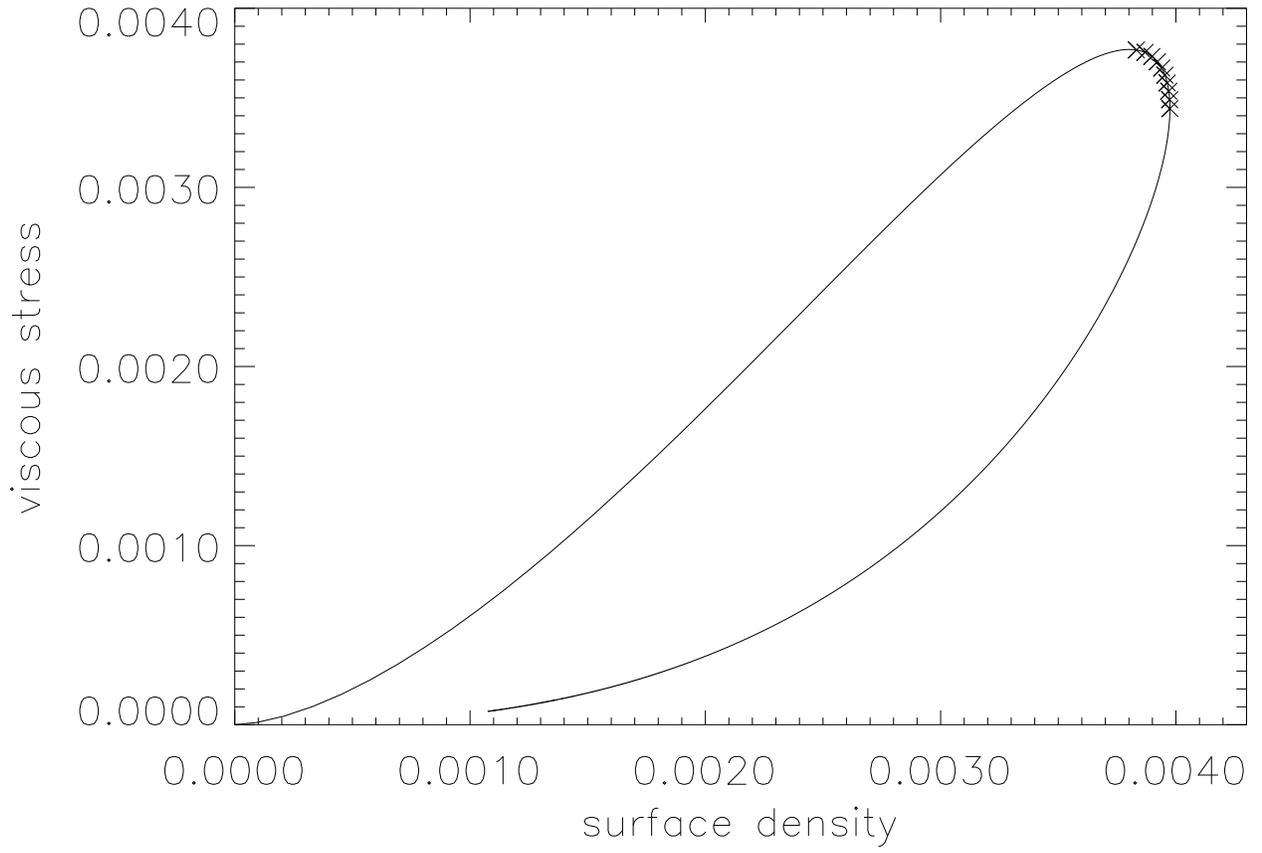
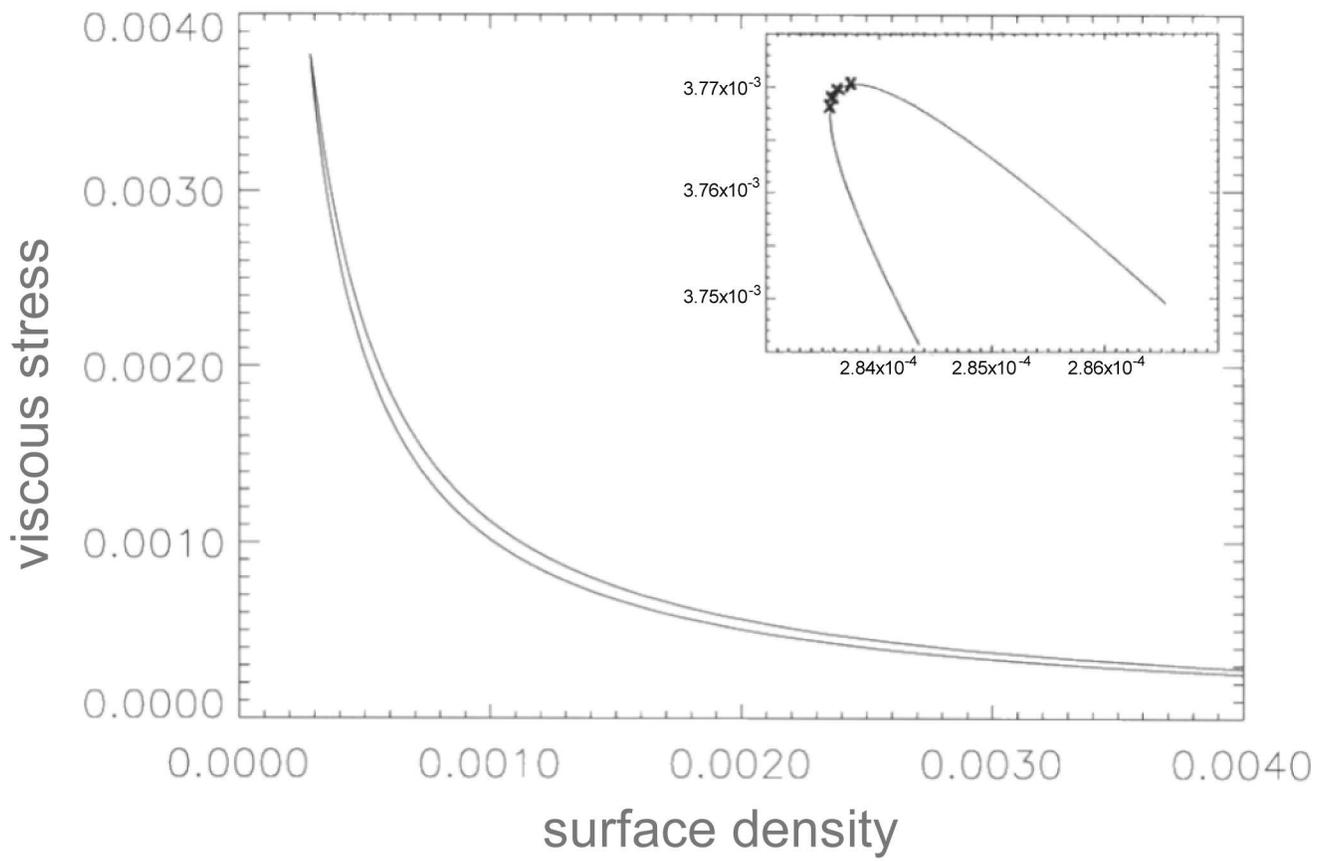
\end{document}